\documentstyle[12pt]{article}

\def\n{\noindent}
\def\half{{1\over 2}}
\def\duom{\dot{{\bf A}}}
\def\uom{{\bf A}}
\def\upsi{{\bf {\Psi}}}
\def\uf{{\bf f}}
\def\be{\begin{equation}}
\def\ee{\end{equation}}
\def\ba{\begin{eqnarray}}
\def\ea{\end{eqnarray}}

\newcommand{\blankline}{\vskip .3cm}
\begin{document}
\begin{titlepage}
\rightline{IUCAA 94/29}
\rightline{September 1994}
\vskip 1.5cm
\centerline{\LARGE The General Self-dual solution of the Einstein Equations}
\vskip 1cm
\centerline{Sucheta Koshti${}^\dagger$ and Naresh Dadhich${}^*$}
\vskip 0.5cm

\centerline{\it Inter-University Center for Astronomy and Astrophysics,}
\vskip2pt
\centerline{\it Pune-411007, India.}
\vskip 1cm
\begin{abstract}

\n We obtain the most general explicit (anti)self-dual solution of the
Einstein equations. We find that any (anti)self-dual   solution can be
characterised by three free functions of which one is harmonic. Any stationary
(anti)self-dual   solution can be characterised by a harmonic function. It
turns out that the form of the Gibbons and Hawking multi-center metrics is the
most general stationary (anti)self-dual solution. We further note that the
stationary (anti)self-dual   Einstein equations can be reinterpreted  as the
(anti)self-dual   Maxwell equations on the Euclidean background metric.

\blankline
\vfill
${}^*
\ $ nkd@iucaa.ernet.in,
$\ \ {}^\dagger$  sucheta@iucaa.ernet.in

\end{abstract}
\end{titlepage}

\section{Introduction}

Real (anti)self-dual   solutions are the  solutions of the Euclidean Einstein
equations (EE) having (anti)self-dual[(A)SD] Riemann tensor. In Quantum gravity
, (A)SD solutions are significant, for, they correspond to saddle points of the
Einstein-Hilbert action, therefore giving large contributions to a path
integral over Euclidean metrics \cite{SWH1}. They can be interpreted as ``one
particle states" in quantum gravity \cite{PENR}. In four dimensions, the
necessary condition for a metric to be hyperk\"ahler is that, it must have
either SD or ASD curvature tensor \cite{ATHISI} .

 There have been several attempts made to tackle the problem of constructing
the general (A)SD solution of the Einstein equation.   Ashtekar \cite{ASHT} has
simplified the self-dual  Einstein equations (SDE) by using the new Hamiltonian
variables for general
relativity. In terms of Ashtekar's  new Hamiltonian variables, SDE may be
rewritten as evolution
equations
for three divergence free vector fields given on a three dimensional
surface
 with a fixed volume element. Penrose \cite{PENR} has used twistor techniques
and shown that , in pricipal, one can construct the general SD solution but the
problem of constructing the Penrose's deformed twistors is  as dificult as
solving the partial differential equations. Grant \cite{GRAN} has shown that
any SD metric can be characterised by a function that satisfies a non-linear
evolution equation , to which the general solution can be found iteratively.
Grant's general SD solution is formal and not explicit. In this paper our main
motivation is to find explicitly, the most general (A)SD solution of the EE.

 (Anti)Self-duality of the Riemann tensor automatically ensures the vanishing
of the Ricci tensor, and so they are solutions of the vacuum EE with vanishing
cosmological constant. In four dimensions, the Riemann tensor is (A)SD iff  the
curvature 2-form $R^a~_b$ is (A)SD.  The (A)SD 2-form $R^a~_b$ can be
considered to come from a (A)SD connection $\omega ^a~_b$ by chosing an
appropriate gauge called ``(A)SD gauge". Therefore one way to find the general
(A)SD solution of the EE is to write the most general form of the  metric and
demand for the (anti)self-duality  of its O(4) Levi-Civita connection. This way
the problem reduces to solving first order partial differential equations. In
this paper we find the general (A)SD solution of the vacuum EE by demanding the
(anti)self-duality  of the O(4) Levi-Civita connection of the most general
metric.

 Here we work in four dimensions and with the Euclidean metric signature, for,
real, nontrivial (A)SD solutions exist only in the Euclidean signature. In
section 2, we write the general form of the metric in terms of six unknown
functions of four variables (t,x,y,z) and show that any (A)SD solution can be
characterised by three free functions  of which one  is harmonic. In section 3,
we show that  any stationary (A)SD solution can be characterised by one free
harmonic function.  It is also demonstrated  that  the stationary (A)SD
solutions of the EE can be reintepreted as  the static (A)SD  Maxwell solutions
on the Euclidean background metric. Section 4 deals with  some particular
(A)SD solutions and  gives a strategy to generate non-stationary (A)SD
solutions from the stationary   (A)SD solutions.   Conclusions are summarised
in  section 6.

\blankline

\section{ Construction of the general (A)SD solution of the vacuum EE.}

In four dimensions, the general metric is characterised by ten functions,
$g_{ij}$'s  of the  four  coordinates. All the ten $g_{ij}$'s are not
independent but they are subject to four coordinate conditions, the Bianchi
identities. Therefore there are only six independent $g_{ij}$. By means of
coordinate transformations, we can arbitrarily assign four of the ten $g_{ij}$,
provided  it does not lead to a  reduction in the dimensions. Therefore without
loss of generality we can assume the form of the most general metric as
follows.

\be
ds^2= u^{-1}(dt+\uom  \cdot  {\bf dx})^2+v_1 dx^2+v_2 (dy^2+ dz^2).
\label{zone}
\ee

\n where $u,v_1,v_2$ and $A_a,  a = 1,2,3$ are functions of t, x, y, z.
\n We chose the vierbein or the tetrad :

\be
e^a~_\mu =\lbrace u^{-\half }(dt+\uom \cdot  {\bf dx}), v_1^\half  dx,
v_2^\half dy, v_2^\half dz\rbrace ,
\ee

\n where $a$ and  $\mu$ are the tetrad  and  spacetime indices respectively
which run from 0 to 3. For the metric (1),  the $O(4)$  Levi-Civita connection
compatible with $e^a$
is given by ,

\ba
\omega ^0~_1 &=& -\half e^0[u^{-1}v_1^{-\half }(u,_x-\dot{u} A_1)+v_1^{-\half
}\dot{A}_1]-\half e^1 [u^\half v^{-1}_1 \dot{v} _1]\nonumber\\
& &+\half e^2 [(uv_1v_2)^{-\half } [(\duom \times {\uom  }+ \nabla \times \uom
)_3]]\nonumber\\
& &-\half e^3 [(uv_2v_1)^{-\half } [(\duom  \times \uom + \nabla  \times  \uom
)_2]]\nonumber\\
& &\nonumber\\
\omega ^2~_3&=&-\half e^0[u^{-\half } {v_2}^{-1} [(\duom  \times \uom +
\nabla  \times  \uom  )_1]]-\half e^2[ v^{-{3\over 2}}_2  (\dot{ v}
_2A_3-v_2,_z)]\nonumber\\
& &-\half e^3[v^{-{3\over 2}}_2 (-\dot{ v} _3A_2+v_2,_y)]\nonumber\\
& &\nonumber\\
\omega ^0~_2 &=& -\half e^0[u^{-1}v_2^{-\half }(u,_y-\dot{u} A_2)+v_2^{-\half
}\dot{A}_2] \nonumber\\
& &-\half e^1[(uv_1v_2)^{-\half } [(\duom \times {\uom  }+ \nabla \times \uom
)_3]]-\half e^2[u^\half v^{-1}_2 \dot{ v} _2 ] \nonumber\\
& &+\half e^3[u^{-\half }{v_2}^{-1} [(\duom  \times \uom
+ \nabla  \times  \uom  )_1]]\label{zom}\\
& &\nonumber\\
\omega ^3~_1 &=& -\half e^0[(uv_2v_1)^{-\half } [(\duom  \times \uom + \nabla
\times  \uom  )_2]]-\half e^1[v^{-1}_1v^{-\half }_2 (-\dot{ v}
_1A_3+v_1,_z)]\nonumber\\
& &-\half e^3[v^{-1}_2v^{-\half }_1 (\dot{ v} _3A_1-v_2,_x)]\nonumber\\
& &\nonumber\\
\omega ^0~_3 &=& -\half e^0[u^{-1}v_2^{-\half }(u,_z-\dot{u} A_3)+v_2^{-\half
}\dot{A}_3]\nonumber\\
& &+\half e^1[(uv_2v_1)^{-\half } [(\duom  \times \uom + \nabla  \times  \uom
)_2]]\nonumber\\
& &-\half e^2[u^{-\half }{v_2}^{-1} [(\duom  \times \uom +
\nabla  \times  \uom  )_1]]-\half e^3[ u^\half v^{-1}_2 \dot{ v} _2
]\nonumber\\
& &\nonumber\\
\omega ^1~_2 &=& -\half e^0[(uv_1v_2)^{-\half } [(\duom \times {\uom  }+ \nabla
\times \uom  )_3]]\nonumber\\
& &-\half e^1[v^{-1}_1v^{-\half }_2 (\dot{ v} _1A_2-v_1,_y)]-\half e^2[v^{-1}_2
v^{-\half }_1 (-\dot{ v} _2A_1+v_2,_x)]\nonumber
\ea

\n where over dot ($\dot{~}$) is the partial differentiation with respect to t.

In our conventions , the Riemann tensor is (A)SD iff it satisfies,
\be
R_{abcd} = \mp ~^*R_{abcd}\equiv \mp \half \epsilon _{abfg} R^{fg}~_{cd},
\ee

\n where $*$ is the Hodge-dual operator and $\epsilon_{abfg}$ is the
Levi-Civita tensor.  The Riemann tensor satisfies the cyclic identity $R^a
{}~_{[bcd]}=0$. Therefore  the  (anti)self-duality  of the Riemann tensor
automatically ensures the vanishing of the Ricci tensor, and so they are
solutions of the vacuum Einstein equations with vanishing cosmological
constant. In four dimensions, the Hodge duality takes 2-forms to 2-forms. The
most important 2-form associated with a four dimentional metric is its
curvature 2-form $R^a~_b$.  Demanding the (anti)self-duality  of the Riemann
tensor is equivalent to demanding the (anti)self-duality  of $R^a~_b$.  Any
(A)SD 2-form $R^a~_b$ can be considered to come from a (A)SD connection $\omega
^a~_b$ by chosing an appropriate gauge called ``(A)SD gauge".
Suppose $R^a~_b$ is (A)SD but  $\omega ^a~_b$ is not (A)SD. Then by decomposing
$\omega ^a~_b$ into SD and ASD parts and using an $O(4)$ gauge transformation,
one can always remove the (A)SD part leaving $\omega ^a~_b$ SD or ASD. The only
change in $R ^a~_b$ under the gauge transformation is a rotation by an
orthogonal matrix which does not change its duality. The metric (\ref{zone})
will be (A)SD if $\omega^a~_b$ given by (\ref{zom} ) is (A)SD. By demanding for
the (anti)self-duality of the connection given by (\ref{zom} ), we get,

\be
\dot{v_1} =\dot{v_2} = 0,
\label{zdv}
\ee
\n and
\be
v_1 = e^{h(x)} \,  v_2,
\label{zv}
\ee

\n where $h(x)$ is an integration  function.  The function $h(x)$ can be
absorbed by redefining the coordinates and without loss of generality one can
assume $v_1= v_2 \equiv v(x,y,z)$. Then the   metric (\ref{zone})  reduces to :

\be
ds^2= u^{-1}(dt+\uom \cdot {\bf dx})^2+ v (dx^2+ dy^2 + dz^2).
\label{zthree}
\ee

\n Substituting $v_1= v_2 \equiv v(x,y,z)$ in (\ref{zom}), (anti)self-duality
conditions on  the connection $\omega ^a~_b$, further leads to  ,

\be
\duom +\Phi \uom -\upsi = 0,
\label{zomd}
\ee

\n and

\be
\duom  \times \uom + \nabla  \times  \uom = \mp 2 u^\half \nabla v^\half ,
\label{zcurlom}
\ee

\n where $\Phi$ and $\upsi$ are given by,

\be
\Phi = \partial_t(\ln{u^{-\half}})
\ee

\n and

\be
\upsi = \nabla ( {\ln{v^\half \over{u^\half}}}) .
\label{zpsi}
\ee

\n Substituting the vector equation (\ref{zomd}) in the vector equation
(\ref{zcurlom}) we get,
\be
\upsi \times \uom + \nabla  \times  \uom = \mp 2 u^\half \nabla v^\half .
\label{zpsicr}
\ee
\n Solving the equations (\ref{zomd}) and (\ref{zcurlom}) is equivalent to
solving the equations (\ref{zomd}) and (\ref{zpsicr}). Therefore  solving the
equations
(\ref{zomd}) and (\ref{zpsicr}) simultaneously, we get,

\be
\uom = u^{\half}[\int_0^t\,( u^{-\half} \upsi  )\,dt\,+\,\uf (x,y,z)],
\label{zuom}
\ee
\n where $\uf$ satisfies the equation :
\be
\nabla \times (v^\half \uf ) = \mp \nabla v.
\label{zuf}
\ee

\n  The above equation  implies that , $v$ is a harmonic function.   Note that,
 given a harmonic function $v$ , there are many solutions  of the equation
(\ref{zuf}). Given $v$, if $\uf$ is a solution ,  then $\uf + v^{-\half} \nabla
\phi $ for an arbitrary $\phi $, is also a solution of the equation
(\ref{zuf}). This implies , given $u$ and $v$, if $\uom$ is a solution of the
equations (\ref{zomd}) and (\ref{zpsicr}), so is $\uom  + u^{\half} v^{-\half}
\nabla \phi $,  for an arbitrary function $\phi $.   The  solution of the
equation (\ref{zuf})  will be unique after fixing the boundry conditions and
the gauge. Here, $\phi$ is a free function and the transformation, $\uom
\rightarrow  \uom  + u^{\half} v^{-\half} \nabla \phi $ leads to  different
(A)SD metrics. This implies that every (A)SD metric is characterised by two
free functions $u$ and $\phi$  and a harmonic function $v$. Given a harmonic
function $v$, the  solution ( see Appendix A) of (\ref{zuf}) is unique up to!
  a gauge and is given by,
\ba
f_x &=& \mp v^{-\half}[\int^{z}_{z_0} v,_y dz \, + \, \partial_x
\phi,]\nonumber\\
f_y &=& \mp v^{-\half}[-\int^{z}_{z_0} v,_x dz \,+ \,
\int^x_{x_0}{v,_z(x,y,z_0)} dx + \, \partial_y \phi]\label{zff}\\
f_z &=& \mp v^{-\half} [\partial_z \phi],\nonumber
\ea
\n where $\phi$ is an arbitrary function.

Thus the metric:

\be
ds^2= u^{-1}(dt+\uom \cdot {\bf dx})^2+ v (dx^2+ dy^2 + dz^2),\nonumber
\ee

\n is the general (A)SD solution of the EE where $u$ and $v$ are free functions
of which $v$ is harmonic and $\uom $  is given by
(\ref{zuom}), (\ref{zpsi})  and (\ref{zff}). Thus any (A)SD metric can be
written in the above form.

\section{The general stationary (A)SD metric.}

The metric (\ref{zthree}) will be general {\it{stationary}} (A)SD metric if $u,
v$ and $\uom$ are independent of t and obey the relations (\ref{zuom}) and
(\ref{zuf}). This implies, $\Psi = 0$ or $u = k v$ where $k$ is an arbitrary
positive constant  and $\uom$ satisfies the relation ,
\be
\nabla \times \uom  = \mp \nabla v.
\label{zsom}
\ee
\n (Note that, a   method of solving the above equation is given in the
Appendix A.)   Without loss of generality, one can assume  $u =v$ , for, by
redefining the coordinates,  one can absorb the constant $k$. Conversely,  if
$u = v$ then by virtue of  (\ref{zpsi}), (\ref{zuom}) and (\ref{zuf}), $\upsi =
0$  and $\uom = v^\half \uf$,  where, $\uf$ is given by the equation
(\ref{zff}). Thus the general stationary (A)SD metric is given by
\be
ds^2= v^{-1}(dt+\uom \cdot {\bf dx})^2+ v (dx^2+ dy^2 + dz^2).\nonumber
\ee
\n where  $\uom = v^\half \uf$, $\uf$ is given by the equation (\ref{zff}) and
$v$ is a free harmonic function.  This implies that, any {\it{stationary}}
(A)SD solution of the EE can be characterised by a harmonic function.  Note
that,  here, the  gauge freedom $\uom \rightarrow \uom + \nabla \phi $ does not
generate a new statinary (A)SD metric but leads to  a  coordinate transformed
(A)SD metric.
\blankline

\n   We notice that the  equation (\ref{zsom}) can be reinterpreted as  the
static (A)SD Maxwell equation on the Euclidean background metric by
reintepreting $\uom$ as the Maxwell vector potential and $v$ as the Maxwell
scalar potential. Thus there is a one-to-one correspondence between
{\it{stationary}} (A)SD solutions of the EE and the static (A)SD Maxwell
solutions on the Euclidean background.

\section{Some particular solutions.}

\blankline
\n The multi-center (A)SD metrics given by Gibbons and Hawking \cite{HAWK2} are
the stationary (A)SD metrics and is given by,

\be
ds^2= v^{-1}(dt+\uom \cdot {\bf dx})^2+ v (dx^2+ dy^2 + dz^2).\nonumber
\ee
\n where,
\be
v = l + 2m \sum_{i=1}^{k}{1\over{\left| \underline{x}-\underline{x_i}\right|}}.
\ee

\n Here,  $l$ and $m$ are constants and $\uom$ can be obtained (see Appendix A)
from the equation,
\be
\nabla \times \uom  = \mp \nabla v.\nonumber
\ee
\n Thus for arbitrary harmonic function $v$, the form of the Gibbons and
Hawking multi-center metric is the most general stationary (A)SD metric.
\blankline

\n  We now give a strategy to generate  non-stationary (A)SD solutions from the
stationary  solutions.

\n {\bf Strategy:} Start with any harmonic function $v(x,y,z)$. Obtain $\uom_s$
by solving (see Appendix A) the equation $\nabla \times \uom_s  = \nabla v.$
This is a stationary (A)SD solution of the EE. Substitute $\uom_s$ in
\be
\uom = u^{\half}[\int_0^t\,( u^{-\half} \upsi  )\,dt\,+\, v^{-\half}
\uom_s],\nonumber
\ee

\n where $u$ is a free function of four coordinates such that $u \not= k v $
and $\upsi = \nabla ( {\ln{v^\half \over{u^\half}}})$. Then the metric:

\be
ds^2= u^{-1}(dt+\uom \cdot {\bf dx})^2+ v (dx^2+ dy^2 + dz^2).\nonumber
\ee

\n is a non-stationary (A)SD solution of the EE.
\section{Conclusions}

\n We have explicitly obtained the most general (A)SD solution of the EE. We
have shown that any (A)SD solution of the vacuum EE can be characterised by
{\it{three}} free functions $u$, $\phi$ and $v$ where $v$ is harmonic.  The
metric is stationary , iff
 $u = k v$. Any {\it{stationary}} (A)SD metric can be characterised by
{\it{one}} free harmonic function. Therefore the problem of finding
{\it{stationary}} (A)SD solutions is equivalent to solving the 3-d Laplace
equation. We also point out that, any {\it{stationary}} (A)SD metric can be
reinterpreted as a {\it{static}} (A)SD Maxwell's solution. The form of the
Gibbons and Hawking multi-center metric is the most general  {\it{stationary}}
(A)SD metric.

\n It should be emphasised that we have not imposed any kind of boundary
conditions on our solutions at infinity. However, one can obtain large variety
of gravitational instantons by imposing appropriate boundary conditions.
\section*{\bf Acknowledgements}

We would like to thank Tarun Souradeep and  Ravi Kulkarni, for valuable
discussions.

 SK  was supported  during this work , by  the National Board for Higher
Mathematics, India, through the Post-doctoral fellowship.
\vfill
\eject

\appendix
\section{The general solution of $\nabla \times \uom = \nabla v $ for given
$v$}

Given $v$, the solution of $\nabla \times \uom = \nabla v $ will be unique upto
a gauge. Any two solutions ${\uom}_1$ and ${\uom}_2$ will be related by,
\be
\uom_1 = \uom_2 + \nabla \phi,
\ee
\n where $\phi$ is an arbitrary function. Therefore it is sufficient to find a
particular solution of  $\nabla \times \uom = \nabla v $  and all other
solutions can be found by gauge transformation.  Without loss of generality,
let us assume that the coordinates have been chosen such that $\uom$  is
parallel to the xy plane. i.e.
\be
A_z = 0 .
\label{omegaz}
\ee
 The  equation $\nabla \times \uom = \nabla v $  with $A_z = 0 $ implies,
\ba
\label{aone}
- A_y,_z& = &v,_x\\
\label{atwo}
A_x,_z& = &v,_y\\
A_y,_x-A_x,_y & = &v,_z.
\label{athree}
\ea

\n Integrating (\ref{aone}) and (\ref{atwo}) with respect to z, we get,
\be
A_x = \int^{z}_{z_0} v,y dz \, + \, g_1(x,y)
\label{omegax}
\ee
\n and

\be
A_y = -\int^{z}_{z_0} v,x dz \, + \, g_2(x,y),
\label{omegay}
\ee
\n where $g_1 $ and $g_2$ are the integration functions of x and y.
Substituting the above two equations in (\ref{athree}) and using the fact that
$v$ is harmonic,  we get,
\be
v,_z(x,y,z) - v,_z(x,y,z_0) - g_1,_y(x,y) + g_2,_x(x,y) = v,_z(x,y,z).
\ee
\n This implies ,
\be
-g_1,_y(x,y) + g_2,_x(x,y) =  v,_z(x,y,z_0).
\label{afour}
\ee

As a particular solution, we choose $g_1(x,y)=0$. Then the equation
(\ref{afour}) gives, $g_2=\int^x_{x_0}{v,_z(x,y,z_0) \, dx}$.

\n Thus given the harmonic function $v$, The general solution of  $\nabla
\times \uom = \nabla v $ is:
\ba
A_x &=& \int^{z}_{z_0} v,_y dz \, + \, \partial_x \phi,\nonumber\\
A_y &=& -\int^{z}_{z_0} v,_x dz \,+ \, \int^x_{x_0}{v,_z(x,y,z_0)} dx + \,
\partial_y \phi\nonumber\\
A_z &=&  \partial_z \phi,\nonumber
\ea

\n where $\phi$ is an arbitrary function.

\vfill
\eject

\end{document}